\documentclass[sigconf]{acmart}

\usepackage{booktabs}
\usepackage{multirow}
\usepackage{graphicx}
\graphicspath{{sigspatial2026_submission/}{img/}}
\usepackage{array}
\usepackage{soul}
\usepackage{tikz}
\usetikzlibrary{positioning,arrows.meta,calc,fit,backgrounds}
\usepackage{fontawesome5}

\makeatletter
\def\@affiliationfont{\small\normalfont}
\makeatother

\acmConference[SIGSPATIAL '26]{The 34th ACM International Conference on Advances in Geographic Information Systems}{November 3--6, 2026}{Riverside, CA, USA}
\acmYear{2026}
\copyrightyear{2026}
\acmISBN{}
\acmDOI{}

\title[Multimodal Geospatial RAG on I-GUIDE]{Intelligent Multimodal Retrieval and Reasoning for Geospatial Knowledge Discovery on the I-GUIDE Platform}

\author{Yunfan Kang}
\email{yfkang@illinois.edu}
\affiliation{%
  \institution{University of Illinois Urbana-Champaign}
  \city{Urbana}
  \state{Illinois}
  \country{USA}
}

\author{Erick Li}
\email{zhiyuan5@illinois.edu}
\affiliation{%
  \institution{University of Illinois Urbana-Champaign}
  \city{Urbana}
  \state{Illinois}
  \country{USA}
}

\author{Furqan Baig}
\email{fbaig@illinois.edu}
\affiliation{%
  \institution{University of Illinois Urbana-Champaign}
  \city{Urbana}
  \state{Illinois}
  \country{USA}
}

\author{Wei Hu}
\email{weih9@illinois.edu}
\affiliation{%
  \institution{University of Illinois Urbana-Champaign}
  \city{Urbana}
  \state{Illinois}
  \country{USA}
}

\author{Alexander Michels}
\email{alexander.michels@utdallas.edu}
\affiliation{%
  \institution{The University of Texas at Dallas}
  \city{Richardson}
  \state{Texas}
  \country{USA}
}

\author{Anand Padmanabhan}
\email{apadmana@illinois.edu}
\affiliation{%
  \institution{University of Illinois Urbana-Champaign}
  \city{Urbana}
  \state{Illinois}
  \country{USA}
}

\author{Shaowen Wang}
\email{shaowen@illinois.edu}
\affiliation{%
  \institution{University of Illinois Urbana-Champaign}
  \city{Urbana}
  \state{Illinois}
  \country{USA}
}

\begin{document}

\begin{abstract}
Geospatial knowledge discovery increasingly requires search across heterogeneous artifacts: datasets, maps, notebooks, software, publications, and the provenance links among them. Conventional geoportals support metadata and spatial filtering, but they rarely provide semantic retrieval, graph-aware provenance traversal, and conversational synthesis in one integrated  system. 
This paper presents I-GUIDE Smart Search, a production multimodal geospatial retrieval-augmented generation (RAG) system embedded in the I-GUIDE Platform, and reports on its design, deployment, and evaluation.
The system combines production-maintained OpenSearch keyword, vector, and spatial indexes with a Neo4j knowledge graph and an iterative RAG pipeline for memory-aware query augmentation, reasoning, retrieval-method routing, relevance grading, grounded generation, hallucination and relevance checking. In a single-A100 RAG deployment, I-GUIDE Smart Search supports interactive use up to about 100 concurrent simulated users, reaching 4.4 requests per second with p50 latency near 25 seconds despite 20--50 LLM calls per query. 
For answer quality, we evaluate a four-category benchmark of 170 unique human-filtered user-facing queries, together with ten intent-specific probe sets generated from the deployed indexes and graph. Smart Search improves retrieved evidence coverage and judged answer quality over non-retrieval and naive-RAG baselines, with the clearest gains on exact-identifier, spatially constrained, simple-recommendation, and domain-specific factual queries requiring current indexed evidence.
A controlled diagonal ablation isolates each pipeline component on the queries it is designed to serve, showing that keyword and spatial retrieval are load-bearing for exact-identifier and geographic-extent queries, while graph traversal, iterative reasoning, memory, reranking, and validation make more conditional but measurable contributions to evidence coverage, correctness, completeness, and routing diagnostics.
We distill transferable deployment lessons for spatial RAG systems, covering spatial metadata quality, graph provenance, retrieval routing, interface contracts, refusal-aware evaluation, latency--cost tradeoffs, and the role of the user interface in deployed geospatial cyberinfrastructure.
\end{abstract}



\maketitle


\section{Introduction}

Geospatial data science increasingly depends on the ability to discover, connect, and reuse heterogeneous research artifacts. 
A reproducible spatial analysis may require datasets, maps, Jupyter notebooks, code repositories, scientific publications, and evidence about how these resources relate to one another. 
Existing geoportals and data catalogs provide solid foundations for publishing and filtering geospatial resources, yet researchers still often need to manually bridge multiple systems to find data, inspect methods, trace provenance, and assemble executable workflows. This fragmentation is especially costly for GeoAI and spatially grounded scientific workflows, where model quality, spatial generalization, and trust depend on transparent data discovery and source attribution.


The I-GUIDE Platform is a scientific gateway for geospatial data-intensive research and education. Its users discover resources contributed by multiple projects and peer practitioners, connect those resources to executable workflows, and reuse them in reproducible analyses. I-GUIDE Smart Search extends this gateway from catalog search to conversational geospatial knowledge discovery. It exposes heterogeneous knowledge elements---datasets, notebooks, maps, software, publications, and educational resources---through production-maintained indexes and a provenance graph, then uses retrieval-augmented generation (RAG) to answer user questions from indexed platform evidence.

Deploying spatial RAG in this setting raises questions that are not answered by LLM integration alone. The system must index spatial, semantic, and graph evidence together; route natural-language queries across heterogeneous retrieval methods; preserve provenance in generated answers; refuse when evidence is insufficient; and remain responsive when iterative reasoning runs on finite GPU capacity. These requirements make Smart Search different from an offline geospatial QA prototype: its retrieval backends are updated through user-facing platform workflows, its answers must be traceable to platform knowledge elements, and its serving behavior is constrained by the same infrastructure that supports production use.

The central claim from deployment experience is that spatial RAG quality is constrained less by LLM capability alone than by the completeness of spatial metadata, the coverage of graph provenance, the reliability of retrieval routing, and the serving limits of production infrastructure. We therefore treat I-GUIDE Smart Search as both a retrieval system and an operational case study: the evaluation asks which parts of the deployed system improved discovery, which components introduced bottlenecks or failure modes, and which lessons transfer to other geospatial cyberinfrastructure projects.
During the 17-month production deployment from December 2024 to May 2026, the system design evolved in response to observed user queries and successive benchmark rounds.
Each round exposed operational failure modes, motivated targeted revisions, and, once those revisions stabilized in production, contributed to the architecture described and measured in this paper. The evaluation is therefore both a measurement of the deployed design and a record of the process that shaped it.

The paper makes five contributions:

\begin{enumerate}
    \item We report on the design and 17-month production deployment of I-GUIDE Smart Search, a multimodal geospatial RAG system that indexes heterogeneous knowledge elements
jointly across OpenSearch keyword, vector, and spatial indexes and a Neo4j provenance graph.
    \item We present a memory-aware iterative RAG pipeline that classifies query intent to route across retrievers, fuses their results with a modality-aware reciprocal-rank-fusion reranker, rewrites queries from conversational memory, and validates answers for hallucination and relevance.
    \item We show that the entire I-GUIDE Smart Search stack---OpenSearch, the Neo4j graph, the vLLM-hosted generation model, and the iterative RAG pipeline---is reproducible on a single A100 GPU node, and we characterize its serving performance at this scale.
    \item We evaluate answer quality using a four-category benchmark of 170 unique human-filtered user-facing queries,
   and we evaluate component behavior using ten intent-specific probe sets constructed against the deployed indexes and graph.
    \item We distill transferable deployment lessons for spatial RAG infrastructure, covering multimodal retrieval design, production interface contracts, refusal-aware evaluation, GPU serving constraints, and user-interface feedback signals.
\end{enumerate}

\section{Background and Related Work}

\textbf{Geoportals, catalogs, and metadata standards.}
Geospatial cyberinfrastructure has long relied on catalogs and spatial data infrastructures to make distributed resources findable, accessible, interoperable, and reusable \cite{Wilkinson2016FAIR}. Open-source platforms such as CKAN and GeoNode publish datasets with structured metadata, tags, map interfaces, and spatial filters \cite{CKAN2025,Corti2019GeoNode}. Federated repositories such as DataONE and the Earth System Grid Federation extend discovery across distributed environmental and climate resources \cite{Michener2011DataONE,Williams2016ESGF}. Standards such as DCAT, OGC API--Records, and the SpatioTemporal Asset Catalog (STAC) further specify how dataset, service, and spatiotemporal asset metadata should be exposed and exchanged \cite{DCAT2024,OGCAPIRecords2023,STAC2025}. These systems and standards provide the substrate for discovery, but they still leave users to bridge datasets, notebooks, maps, code, publications, and derived products when the relationships among artifacts are incomplete or scattered across systems, or not expressed in searchable form.

\textbf{Semantic and geographic information retrieval.}
Geographic information retrieval has long studied the joint handling of topical and spatial relevance, including the ambiguity of place names, geographic footprints, and spatial intent in natural-language search \cite{Purves2018GIRSurvey}. Embedding-enhanced geoportal search shows that semantic retrieval can improve recall when user vocabulary differs from metadata vocabulary \cite{Mai2020GeoportalSearch}. Large scholarly corpora such as S2ORC show how linking publication metadata, structured full text, citations, figures, and tables enables richer scholarly retrieval \cite{Lo2020S2ORC}. However, dense semantic similarity alone does not reliably encode spatial overlap, temporal constraints, exact identifiers, data lineage, or workflow provenance. Production geospatial search therefore needs hybrid retrieval that combines lexical, vector, spatial-temporal, and relational evidence rather than replacing structured search with embeddings.

\textbf{Geospatial knowledge graphs and provenance.}
Linked-data standards provide complementary mechanisms for representing spatial and provenance relationships. GeoSPARQL defines vocabularies and query functions for spatial RDF data \cite{GeoSPARQL2024}, while PROV-O defines general provenance concepts for entities, activities, and agents \cite{PROVO2013}. Large-scale resources such as KnowWhereGraph demonstrate how spatial, temporal, and thematic entities can be linked for interdisciplinary discovery and geo-enrichment \cite{KnowWhereGraph2025}. Graphs are especially useful for questions such as which notebooks use a dataset, which publications cite derived outputs, or which resources share contributors and workflows. In practice, however, graph retrieval must be combined with content and spatial search because many user questions mix text, location, time, artifact type, and relationship constraints in a single request.

\textbf{RAG, graph RAG, and agentic retrieval.}
Retrieval-augmented generation grounds LLM outputs in retrieved evidence rather than relying only on parametric model knowledge \cite{Lewis2020RAG}. Recent surveys distinguish dense, sparse, hybrid, corrective, and iterative RAG pipelines, emphasizing that retrieval quality, evidence filtering, reranking, and evaluation often dominate final answer quality \cite{Gao2023RAGSurvey}. Self-reflective and corrective RAG systems use retrieval decisions, critique, or evidence grading to improve factuality when retrieved context is weak or irrelevant \cite{Asai2023SelfRAG,Yan2024CRAG}. GraphRAG-style methods add graph structure for multi-hop or corpus-level synthesis \cite{Edge2024GraphRAG}. These directions motivate the agentic components in I-GUIDE Smart Search: query rewriting, retrieval routing, evidence grading, answer validation, and graph-aware retrieval.
The remaining challenge is operational: these
components must work together over heterogeneous geospatial artifacts,
not only over text documents.

\textbf{Spatial RAG and deployed GeoAI systems.}
Recent geospatial RAG systems combine spatial filtering, dense retrieval, graph queries, or GeoSPARQL translation for spatial question answering and geospatial modeling \cite{SpatialRAG2025,GeoQA22024,GDQA2025,GeoGraphRAG2025}. They demonstrate that LLM-based geospatial assistants need retrieval mechanisms that understand location, scale, topology, and domain metadata. At the same time, deployed RAG systems are constrained by serving infrastructure. Systems such as vLLM improve throughput through efficient model serving, but multi-step agentic pipelines can still require many model calls per user query \cite{Kwon2023vLLM}. 
I-GUIDE Smart Search therefore contributes an experience report on a deployed spatial RAG system that integrates production-maintained OpenSearch indexes, a Neo4j provenance graph, agentic retrieval orchestration, and single-GPU serving constraints inside a geospatial scientific gateway.

\section{System Context and Experience Setting}
\label{sec:system-context}


The I-GUIDE Platform is a scientific gateway for geospatial
data-intensive research and education. It enables users to discover,
submit, connect, and reuse knowledge elements contributed by multiple
projects and community members. Platform resources are represented as
first-class knowledge elements---datasets, maps, notebooks, software,
publications, and educational materials---that share a persistent
identifier across OpenSearch, Neo4j, object storage, external URLs, and
user-facing pages.
This setting differs from a standalone geospatial question-answering
prototype in two ways. First, Smart Search operates over
production-maintained indexes rather than a frozen benchmark corpus:
registered users submit and revise knowledge elements through platform
forms, background services update text, vector, spatial, and graph
indexes, and user activity updates usage-derived signals such as
\texttt{click\_count}, and \texttt{bookmarked-by}. Second, the system is part of a shared scientific
gateway, so search must preserve provenance, respect authenticated APIs,
and remain responsive while invoking LLM-based reasoning. The retrieval
ecosystem must therefore remain self-consistent without relying on
offline batch reconstruction, a deployment condition that most academic
RAG benchmarks do not model.

Five deployment requirements shape the system design. First, search must be \emph{transparent}: users and reviewers should be able to inspect metadata, retrieval paths, cited knowledge elements, and answer provenance. Second, the system must \emph{interoperate} with externally hosted artifacts because many contributed resources are referenced rather than stored directly in I-GUIDE. Third, discovery must support \emph{contextual retrieval} beyond keyword matching because user queries commonly combine concepts, locations, resource types, usage signals, and provenance relationships. Fourth, the RAG path must remain \emph{responsive} even when complex queries trigger multiple LLM calls on finite GPU capacity. Fifth, public discovery and authenticated actions must share a common \emph{identity and authorization} layer so that protected resources and user data are not exposed through search responses.

\begin{figure}[t]
    \centering
    \includegraphics[width=\columnwidth]{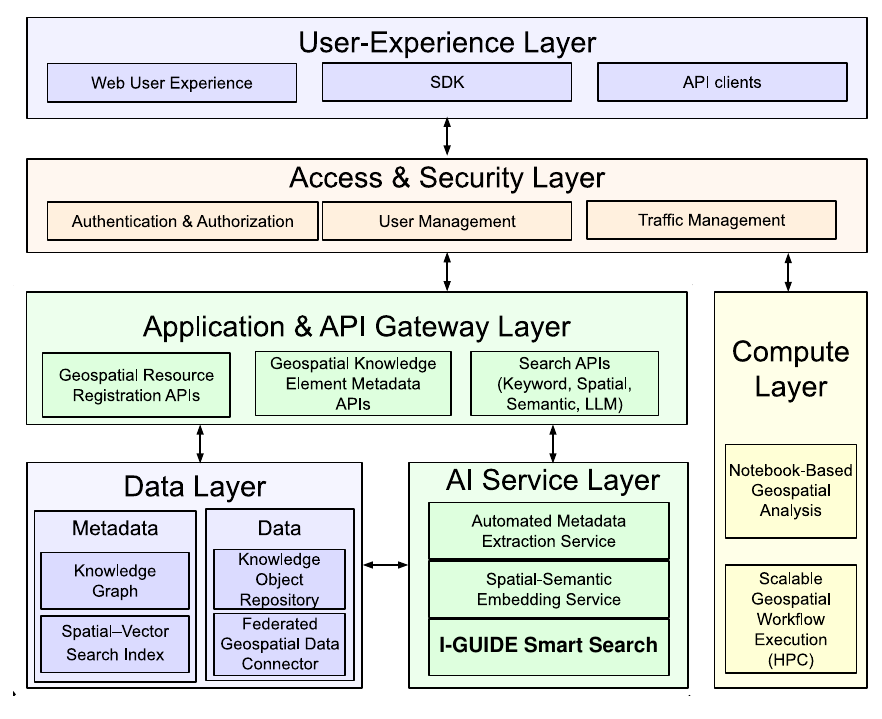}
    \Description{Layered I-GUIDE platform architecture showing user experience, API gateway, security, search and AI services, data storage, and compute resources.}
    \caption{I-GUIDE platform architecture. The search system spans metadata storage, AI services, and API access while relying on compute, security, and user-facing layers.}
    \label{fig:platform}
\end{figure}

The deployment relies on open-source infrastructure, including OpenSearch, Neo4j, vLLM, MinIO, federated connectors, and CILogon for authentication~\cite{OpenSearch2025,Neo4j2026, Qwen2025, Kwon2023vLLM, MinIO2026, Basney2014CILogon}.
In production, Smart Search accesses Qwen2.5-7B-Instruct through AnvilGPT's
vLLM backend~\cite{Anvil25}; in the experiments reported here, we deploy the same model and pipeline ourselves on a single JetStream2 A100 GPU node~\cite{Hancock2021JS2}.
Table~\ref{tab:deployment-snapshot} summarizes the production snapshot used to interpret the benchmarks reported later in the paper. Figure~\ref{fig:platform} situates Smart Search within the broader I-GUIDE Platform architecture. This placement is important for the experience reported here: Smart Search is not a standalone QA service, but a deployed component inside a scientific gateway with shared identity, data-management, indexing, and infrastructure constraints.

\begin{table}[t]
\caption{Production deployment snapshot for I-GUIDE Smart Search as of May 2026.}
\label{tab:deployment-snapshot}
\small
\begin{tabular}{@{}p{0.34\columnwidth}p{0.58\columnwidth}@{}}
\toprule
\textbf{Dimension} & \textbf{Production I-GUIDE Smart Search setting} \\
\midrule
Deployment target & I-GUIDE Platform Smart Search, deployed at \texttt{platform.i-guide.io} \\
\midrule
Indexed artifacts & 697 public knowledge elements: 104 datasets, 160 maps, 167 notebooks, 33 software/code resources, 198 publications, and 35 educational resources as of May 2026 \\
\midrule
Search backends & OpenSearch for keyword, vector, and spatial indexing; Neo4j for graph relationships \\
\midrule
Graph scale & 2,936 Neo4j nodes and 2,416 relationships \\
\midrule
Production LLM serving & Qwen2.5-7B-Instruct accessed through AnvilGPT's vLLM-backed inference service \\
\midrule
Experimental deployment & Full stack reproduced on one Jetstream2 A100 GPU node, including OpenSearch, Neo4j, vLLM, Qwen2.5-7B-Instruct, and the iterative RAG pipeline\\
\midrule
Authentication & CILogon-issued JWT tokens and role-based endpoint access \\
\midrule
Operational period & Production deployment observed from December 2024 to May 2026 \\
\bottomrule
\end{tabular}
\end{table}

\section{Architecture}
\label{sec:architecture}

I-GUIDE Smart Search realizes the five deployment requirements from Section~\ref{sec:system-context} through three connected parts, shown in Figure~\ref{fig:rag}. 
First, the storage and indexing backends represent each knowledge element under a shared identity across OpenSearch and Neo4j, supporting interoperability, provenance tracing, and contextual discovery. 
Second, the agentic query pipeline routes questions across keyword, semantic, spatial, and graph retrieval paths; fuses returned evidence; generates grounded answers; and streams execution progress to the user, supporting both contextual discoverability and transparency.
Third, the self-hosted serving stack keeps LLM reasoning within finite GPU capacity, supports future model replacement and scaling, and makes the full system reproducible under bounded infrastructure resources. 
As part of the AI Service Layer in Figure~\ref{fig:platform}, Smart Search is protected by the \emph{Access \& Security Layer}, so public discovery, registered-user workflows, user data, and LLM services share a single identity and authorization model.

\subsection{Knowledge Representation and Indexing}

Every artifact is represented as a knowledge element with shared metadata fields and type-specific extensions, and one persistent identifier links its OpenSearch
record, Neo4j node, files, external URLs, and user-facing page---so externally hosted resources integrate without forcing uniform storage.

OpenSearch stores the primary searchable representation. Title, authors, tags, descriptions, and other metadata fields are indexed for BM25 keyword retrieval; embedding fields such as the embedded descriptions and pdf-chunks support $k$-nearest-neighbor vector search; and GeoJSON geometries, centroids, bounding boxes, and temporal coverage for spatial-temporal filtering. Publications are chunked for document-level QA, while notebooks and code carry extracted function signatures, dependencies, and inferred analytical tasks. 
The spatial endpoint interprets coordinate inputs as point, envelope, or polygon queries, so one API serves both location lookup and area-based filtering.

Neo4j stores knowledge elements and contributor/user entities as graph nodes, with typed edges capturing their relationships.
Edges encode associations such as contributed-by, authored-by, bookmarked-by, cites, related-to, derived-from, and used-in, enabling questions that are awkward for text search alone, such as which notebooks use a dataset, which publications connect to those notebooks, or which resources a contributor has submitted.
These relationships are exposed both to the RAG pipeline and
to direct browsing APIs.
Neo4j also stores fields and edges labels such as \texttt{click\_count}, which were originally introduced to support frontend sorting by popularity and user bookmarking of knowledge elements, and later became usage-derived signals for popularity-style or graph queries such as ``most viewed'' or ``popular elements.''

Knowledge elements enter the indexes through user-facing registration forms on the I-GUIDE Platform: contributors specify the metadata above and explicit related-knowledge-element edges that populate the Neo4j graph. Extraction services and embedding updates run in background threads triggered by form submission, so the OpenSearch and Neo4j indexes evolve continuously without manual reindexing. 
User interactions such as browsing, search, and card clicks increment \texttt{click\_count} in near real time. The retrieval ecosystem therefore remains production-maintained rather than reconstructed as an offline benchmark corpus.

\subsection{Agentic Query Pipeline}

\begin{figure*}[t]
\centering
\includegraphics[width=\textwidth]{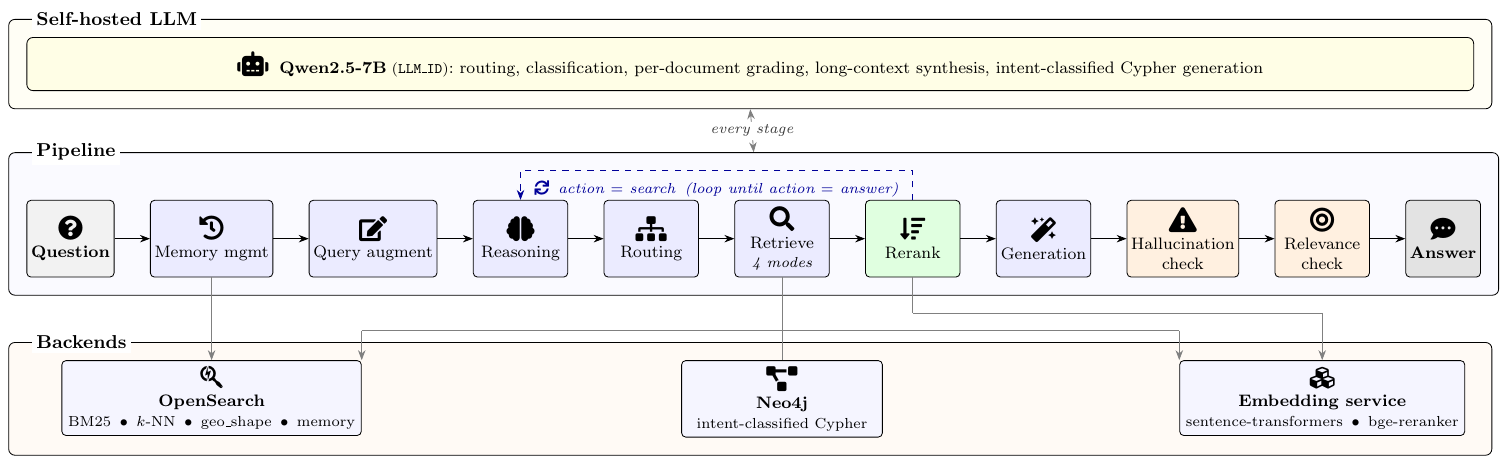}
\Description{Production iterative RAG pipeline rendered as three stacked peer
groups. The top group, ``LLM stack'', is a single self-hosted Qwen2.5-7B model
served with vLLM that every pipeline stage calls for routing, classification,
per-document grading, intent-classified Cypher generation, validation, and
synthesis. The middle group, ``Pipeline'', is a horizontal flow of stages from
Question to Answer with icons, and a dashed iteration loop from Rerank back to
Reasoning. The bottom group, ``Backends'', contains OpenSearch, Neo4j, and the
embedding service. A single double-headed arrow links the Pipeline and LLM-stack
boxes; downward arrows on separate lanes indicate which stages access which
backend.}
\caption{Production I-GUIDE Smart Search RAG pipeline. 
}
\label{fig:rag}
\end{figure*}

Figure~\ref{fig:rag} shows the query pipeline. A \emph{memory} module first checks
whether a new question depends on recent dialogue and, if so, rewrites it with the
needed context.
A \emph{query-augmentation} module then refines the contextualized
question. 
The reasoning module drives a bounded ReAct-style loop: for simple queries it can proceed directly to retrieval or answer generation, while for complex queries it decomposes the request into smaller subqueries, decides which evidence is still missing, and emits a structured action at each step--- \texttt{search} when more evidence is needed or \texttt{answer} when sufficient evidence has accumulated. 
The deployed multi-hop path caps the loop at four iterations so a query cannot reason indefinitely.


When the reasoning module requests search, the routing module selects retrieval methods through a combination of hand-curated trigger patterns and a few-shot LLM router. Triggers augment the LLM selection rather than short-circuit it, so a query carrying both topical and spatial cues can route to keyword, semantic, and spatial paths simultaneously.
The retrieval registry covers OpenSearch keyword retrieval for exact textual matches, semantic vector retrieval for conceptually similar resources, spatial search for location-bearing queries, and Neo4j queries for structural and usage-derived signals that are not captured in document text.
The Neo4j query module uses an \emph{intent-classified Cypher catalog}
of ten templates (Table~\ref{tab:cypher-catalog} in
Appendix~\ref{app:cypher-catalog}): the agent classifies each question into one
structural intent using the LLM and emits Cypher conforming to the matching
template.
This bounded catalog is a deliberate tradeoff. In early deployment, the Qwen2.5-7B-Instruct model was unreliable at open-ended text-to-Cypher generation without examples, but unconstrained few-shot prompting also overfit the model toward the example query shapes. We therefore use intent classification followed by template-constrained Cypher generation: the examples teach the model the schema and output form, while the intent catalog bounds the space of valid graph queries. Each template returns a standard projection, so downstream stages see a uniform document shape regardless of retrieval path.

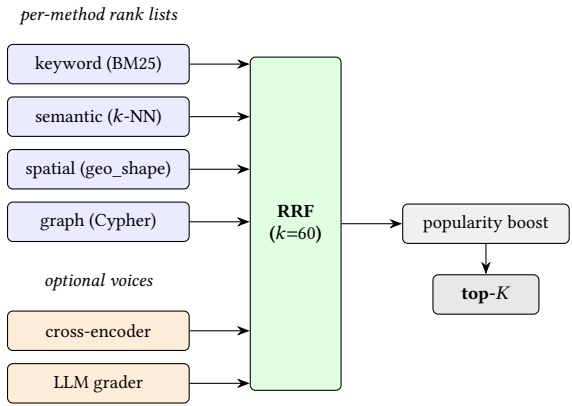
\begin{figure}[t]
\centering
\begin{tikzpicture}[
    node distance=1.6mm,
    box/.style   = {draw, rounded corners=2pt, font=\footnotesize,
                    minimum height=5.2mm, minimum width=24mm,
                    align=center, inner sep=2pt},
    retr/.style  = {box, fill=blue!8},
    voice/.style = {box, fill=orange!15},
    fuse/.style  = {box, fill=green!12, font=\bfseries\footnotesize,
                    minimum width=12mm, align=center},
    post/.style  = {box, fill=gray!12, font=\footnotesize, minimum width=22mm},
    outbox/.style= {box, fill=gray!18, font=\bfseries\footnotesize,
                    minimum width=14mm},
    grp/.style   = {font=\footnotesize\itshape},
    a/.style     = {->,>=Stealth, thin}
]
\node[grp]                          (lab_r) {per-method rank lists};
\node[retr, below=1.5mm of lab_r]   (kw)    {keyword (BM25)};
\node[retr, below=of kw]            (sem)   {semantic ($k$-NN)};
\node[retr, below=of sem]           (sp)    {spatial (geo\_shape)};
\node[retr, below=of sp]            (gr)    {graph (Cypher)};
\node[grp, below=3mm of gr]         (lab_v) {optional voices};
\node[voice, below=1.5mm of lab_v]  (ce)    {cross-encoder};
\node[voice, below=of ce]           (llm)   {LLM grader};
\node[fuse, minimum height=44mm] (rrf)
    at ($(kw.east)!0.5!(llm.east) + (14mm, 0)$) {RRF\\($k{=}60$)};
\node[post, right=8mm of rrf] (boost) {popularity boost};
\node[outbox, below=4mm of boost] (top) {top-$K$};
\foreach \src in {kw, sem, sp, gr, ce, llm}
  \draw[a] (\src.east) -- (\src.east -| rrf.west);
\draw[a] (rrf.east) -- (boost.west);
\draw[a] (boost.south) -- (top.north);
\end{tikzpicture}
\caption{RRF-based rank fusion in the rerank module.}
\label{fig:rerank}
\end{figure}

The \emph{retrieval} stage returns each method's ranked candidate list,
preserving the per-method rank rather than collapsing candidates into a single
score. The \emph{rerank} stage fuses these lists via Reciprocal Rank Fusion
(RRF)~\cite{Cormack2009RRF} with $k=60$, augmented by optional voices from a
cross-encoder and a pointwise LLM grader. RRF is
scale-free, which is essential when combining BM25 scores, vector similarities, spatial intersection signal, and graph-derived
evidence in one
ranking as illustrated in Figure~\ref{fig:rerank}. A small multiplicative log-popularity boost
from \texttt{click\_count} breaks ties without overriding relevance.

Once the reasoning module decides the accumulated evidence is sufficient, the
\emph{generation} module synthesizes a concise answer grounded in the top-ranked
elements. Two \emph{validation} modules then check whether the answer is supported by the retrieved evidence and relevance to the original question.
Failed validation triggers up to three bounded
regeneration attempts.
If the system still cannot produce a supported answer, it emits an explicit insufficient-information response rather than guessing.
Smart Search does not use web-search or parametric-knowledge fallback: it is deliberately scoped to the indexed I-GUIDE knowledge base so every claim in a generated answer can be traced to a user-facing knowledge-element card.

Throughout the pipeline, Server-Sent Events (SSE) stream progress during multi-step execution, reporting stages such as question augmentation, iterative retrieval, generation, memory update, final answer construction, returned reference documents, and errors that prevent successful answering. 
This streamed trace makes the pipeline's intermediate steps inspectable, supports end-to-end transparency, and reduces perceived latency during the 5--30 second
wall-clock time typical of multi-step retrieval.
Each RAG interaction also exposes an optional user-feedback form, with thumbs-up/down and per-dimension ratings for relevance, sufficiency, accuracy, clarity, completeness, and trust; these ratings are stored with the interaction round for future pipeline monitoring and improvement, but are not used as evaluation labels in this paper.

\subsection{Serving and Infrastructure}

For the experiments reported in this paper, we reproduce the full stack on
one JetStream2 virtual machine with a single NVIDIA A100 GPU and serve
Qwen2.5-7B-Instruct through vLLM for all LLM-side stages---routing, classification, per-document grading, Cypher
generation, validation, and synthesis---so that the complete system, including
OpenSearch and Neo4j, can run on one JetStream2 virtual machine with a single NVIDIA A100 GPU. 
The pipeline is model-agnostic per stage, so a larger model can be configured for synthesis or Cypher generation
when additional GPU capacity is available.
We report the single-model
configuration because it is the one we benchmarked and the one reproducible on a single GPU. 
Because a complex query can issue 20--50 LLM calls across reasoning,
routing, grading, generation, and validation, the production configuration
exposes a standard retrieval path alongside the iterative multi-hop path, so
expensive reasoning is reserved for complex questions rather than applied to every
search.


\section{Experimental Evaluation}
\label{sec:eval}

We evaluate I-GUIDE Smart Search as a deployed RAG system rather than as a standalone LLM.
Accordingly, Section~\ref{sec:eval} is organized as an experience-style evaluation: each measurement is tied to an operational decision we had to make during deployment, rather than treated as a generic benchmark of model ability. The evaluation asks five questions: (1) whether the single-GPU experimental stack sustains the interactive concurrent load expected of a small- to medium-sized scientific gateway; (2) whether the production pipeline
produces better-grounded and higher-quality answers than non-retrieval LLMs and a semantic-only RAG baseline; (3) how a deployed
RAG pipeline built around a small open model compares with stronger
frontier non-retrieval LLMs; (4) which pipeline components are load-bearing on which query types, and how their removal changes pipeline behavior; and (5) which deployment failures motivated the revised architecture.

Because Smart Search is a platform-grounded discovery system, the evaluation is constructed against the I-GUIDE knowledge base rather than a general geospatial QA benchmark. The goal is not to demonstrate universal chatbot superiority, but to test whether Smart Search can answer discovery questions that fall within the topical scope of the I-GUIDE Platform using traceable evidence from current platform indexes. This directly addresses why users cannot simply rely on a general-purpose chatbot: in practice, the I-GUIDE knowledge base changes as datasets, notebooks, maps, contributors, and graph links are added or revised, while an LLM's parametric knowledge may be partial, stale, or unreliable for these platform-specific and domain-specific facts.

\subsection{RAG Serving Benchmark}
\label{sec:serving}

We benchmarked the experimental deployment with Locust against a Jetstream2 virtual machine with one NVIDIA A100 GPU.
This deployment reproduces the full Smart Search stack used for the paper experiments: OpenSearch, Neo4j, vLLM, the same Qwen2.5-7B-Instruct model used in production, and the iterative RAG pipeline.
The benchmark OpenSearch and Neo4j databases store a snapshot of the production indices and graph.
Test queries were sampled from frequent platform query patterns, and each request exercised the deployed pipeline stages for memory handling, reasoning, routing, grading, generation, and validation. In observed runs, complex queries required 20--50 LLM calls.
Simulated users waited 5--10 seconds between requests, and for each load level we recorded throughput and latency over a stable one-minute interval.

\begin{figure*}[t]
    \centering
    \begin{minipage}{0.48\textwidth}
        \centering
        \includegraphics[width=\linewidth]{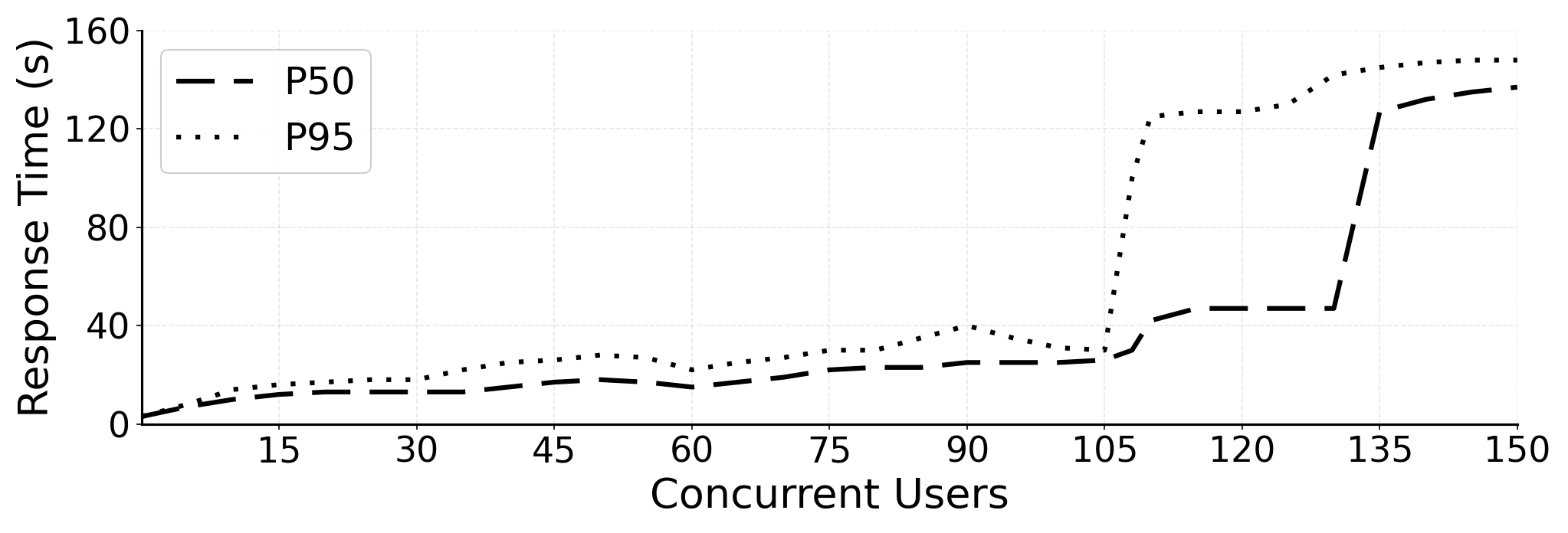}
        \Description{Line chart showing RAG response time increasing as concurrent users increase.}
        \centerline{\small (a) Response time}
    \end{minipage}
    \hfill
    \begin{minipage}{0.48\textwidth}
        \centering
        \includegraphics[width=\linewidth]{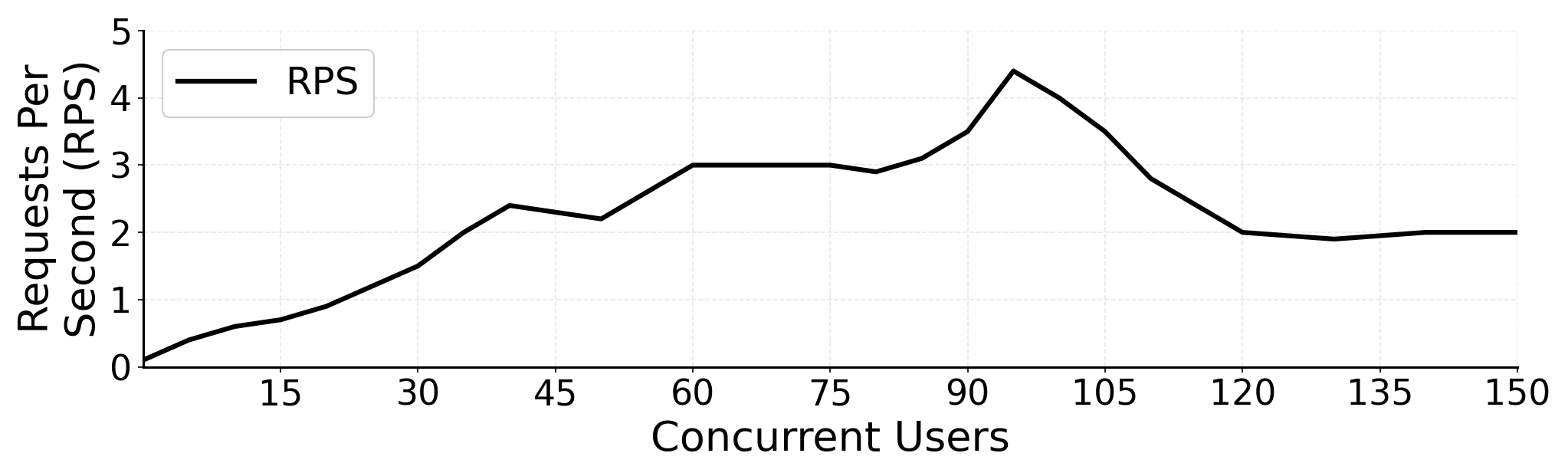}
        \Description{Line chart showing RAG requests per second increasing up to a saturation point and then declining at high concurrency.}
        \vspace{-0.8em}
        \centerline{\small (b) Requests per second}
    \end{minipage}
    \caption{RAG pipeline latency and throughput under increasing concurrent users.}
    \label{fig:rag-eval}
\end{figure*}
Figure~\ref{fig:rag-eval} shows that a single user sees an average response time of 4.9 seconds. At 100 concurrent users, the system reaches a stable 4.4 requests per second, with p50 latency around 25 seconds and p95 latency around 30 seconds. Beyond roughly 130 concurrent users, throughput declines and latency increases, indicating that requests begin to queue behind GPU-bound LLM inference. We treat this as an operational boundary, not as a universal capacity claim: the result identifies the point at which the current single-GPU, single-model deployment must either shed optional reasoning work or scale horizontally.

The deployment lesson is direct: iterative RAG is viable for a small- to medium-sized scientific gateway, but finite GPU capacity shapes both model choice and pipeline design. Serving a small open model keeps the system deployable on bounded infrastructure, but the smaller context window favors pointwise operations for stages such as per-document relevance grading. This reduces prompt size but increases the number of LLM calls per complex query. The experience lesson is therefore a routing lesson: the production system should reserve multi-step reasoning, grading, and validation for queries whose intent requires them, while allowing simpler catalog lookups to take a shorter path.

\subsection{Answer Quality Benchmark}
\label{sec:answer-quality-benchmark}


\paragraph{Benchmark datasets}
The benchmark was built to reproduce the kinds of evidence failures we saw while operating Smart Search, not to define a general-purpose geospatial QA leaderboard.
We retained 532 elaborative QA instances from the I-GUIDE knowledge base, grouped into two benchmark cohorts:

\begin{itemize}
\item \textbf{Main user-facing benchmark} (270 instances): a canonical subset of 170 questions manually filtered from 200 generated candidates across four user-facing categories---simple recommendation, simple direct-answer, complex multi-hop recommendation, and complex multi-hop direct-answer---plus 100 regenerated simple-query variants for paraphrase robustness.
\item \textbf{Targeted ablation benchmark} (262 rows across ten intent-specific probe sets): each set is constructed to require one specific retrieval or generation behavior, and pairs with the corresponding diagonal ablation in Section~\ref{sec:ablations}.
\end{itemize}

The candidates were drafted bottom-up from I-GUIDE metadata, document chunks, and graph edges across two generation waves. This bottom-up construction keeps the evaluation tied to records, spatial footprints, and graph edges that the deployed system could actually retrieve. The 170 canonical user-facing questions were first drafted with \texttt{o1-2024-12-17} in May 2025 to test the behavior of the production pipeline, then revised against the May 2026 database snapshot. The 100 regenerated simple-query variants and the 262 targeted ablation rows were drafted with \texttt{claude-opus-4.7} in May 2026. Each instance records a question, reference answer, source knowledge-element identifiers, expected query type, and complexity label. Human reviewers checked the canonical user-facing candidates and retained 170 from the original 200 generated questions; the regenerated variants and ablation probes were retained as targeted evaluation sets. We report the paraphrase robustness results separately in Appendix~\ref{app:paraphrase}.

\paragraph{Methods compared.}
We compare six methods. \textbf{Qwen2.5} (\texttt{qwen2.5\\-7b-instruct}) is 
prompted directly without retrieval;
it is also the generation backbone of both Naive RAG and the I-GUIDE
Smart Search pipeline.
\textbf{GPT-4o} (\texttt{gpt-4o-2024-11-20}) is a stronger
non-retrieval model and represents the advanced-model class at the time Smart Search was originally deployed.
\textbf{GPT-5.5} (\texttt{gpt-5.5-2026-04-23} via OpenAI API, per-question with
\texttt{max\_completion\_tokens} 8000 to accommodate hidden reasoning
tokens) is a frontier reasoning-model non-retrieval baseline, and
\textbf{Opus 4.7} (\texttt{claude-opus-4.7} via Claude Code CLI) is a frontier non-retrieval baseline paired with a modern coding agent framework.
These two models represent the current state of the art among publicly available LLMs; we include them to test how much factual ground recent frontier models already cover and whether the deployed multimodal RAG pipeline retains its value against them.
The four non-retrieval baselines see only the original system prompt
(``You are a geospatial expert assistant answering factual questions clearly and concisely.'') and the question text.
No external search, file access, or tool execution was enabled for the non-retrieval baselines.
\textbf{Naive RAG} performs single-pass semantic retrieval with
Qwen2.5 generation and no graph, spatial, keyword, or agent
retriever, no grader, and no validator.
\textbf{I-GUIDE Smart Search} is the deployed multimodal pipeline.
All methods used aligned answer-format instructions to reduce formatting effects.

\paragraph{Metrics.}
We report four metrics. \textbf{Recall@10} measures whether the top ten retrieved knowledge-element identifiers include the reference
elements. It therefore applies only to retrieval methods. 
\textbf{Faithfulness} measures answer honesty: an honest abstention is faithful, whereas fabricated facts---dates, authors, provenance, or resource identifiers---are not, even when the resulting answer is correct. \textbf{Correctness} measures whether the answer addresses the asked question and is accurate, crediting correct content beyond the reference. \textbf{Completeness} measures how comprehensively the answer addresses the question. The three axes are scored independently by a blind \texttt{gpt-5-mini} LLM-as-judge that sees only the question, reference answer, and candidate answer, not the method label. Scores are reported on a 0--100 scale.

{\paragraph{Human calibration.} To check that the judge is not merely validating itself, two authors independently scored a stratified audit sample of 50 anonymized answer instances
(Appendix~\ref{app:human-calib}). Judge--human agreement on correctness and completeness ($\kappa$ 0.52 and 0.69; Spearman 0.61 and 0.68) met or exceeded inter-annotator agreement ($\kappa\approx0.30$), indicating that the judge is useful for rank-order comparison at this scale; faithfulness agreement was lower and is therefore interpreted only together with correctness, completeness, and retrieval coverage.

\label{sec:main-bench}

\begin{table*}[t]
\centering
\caption{Main answer quality benchmark on the 170 unique canonical user-facing queries.}
\label{tab:main-bench}
\small
\setlength{\tabcolsep}{2.8pt}
\begin{tabular}{l cccc cccc cccc cccc cccc}
\toprule
\multirow{2}{*}{\textbf{Method}}
& \multicolumn{4}{c}{\textbf{Overall (n=170)}}
& \multicolumn{4}{c}{\textbf{Simple Rec.\ (n=50)}}
& \multicolumn{4}{c}{\textbf{Simple Answer (n=44)}}
& \multicolumn{4}{c}{\textbf{Complex Rec.\ (n=30)}}
& \multicolumn{4}{c}{\textbf{Complex Answer (n=46)}} \\
\cmidrule(lr){2-5} \cmidrule(lr){6-9} \cmidrule(lr){10-13} \cmidrule(lr){14-17} \cmidrule(lr){18-21}
& R@10 & F & C & Cp
& R@10 & F & C & Cp
& R@10 & F & C & Cp
& R@10 & F & C & Cp
& R@10 & F & C & Cp \\
\midrule
Qwen2.5                & -- & 69 & 39 & 51 & -- & 94 & 37 & 44 & -- & 62 & 41 & 57 & -- & 58 & 34 & 50 & -- & 57 & 41 & 56 \\
GPT-4o                 & -- & 58 & 44 & 61 & -- & 95 & 48 & 50 & -- & 20 & 27 & 82 & -- & 52 & 47 & 52 & -- & 58 & 53 & 61 \\
GPT-5.5                & -- & 64 & 55 &\textbf{72}& -- &\textbf{96}& 60 & 57 & -- & 20 & 26 & 85 & -- & 67 & 66 &\textbf{70}& -- & 69 & 68 &\textbf{77}\\
Claude Opus 4.7        & -- & 76 & 56 & 50 & -- & 66 & 43 & 35 & -- &\textbf{97}& 45 & 33 & -- & 68 &\textbf{72}& 69 & -- & 72 &\textbf{70}& 71 \\
Naive RAG              & 38 &\textbf{78}& 41 & 41 & 17 & 67 & 30 & 23 & 72 & 87 & 72 & 77 & 23 & 66 & 30 & 31 & 33 &\textbf{88}& 31 & 31 \\
\textbf{I-GUIDE SS}    &\textbf{73}& 77 &\textbf{72}& 71 &\textbf{63}& 75 &\textbf{76}&\textbf{70}&\textbf{98}& 81 &\textbf{91}&\textbf{97}&\textbf{63}&\textbf{72}& 64 & 64 &\textbf{63}& 80 & 54 & 53 \\
\bottomrule
\end{tabular}
\end{table*}

\paragraph{Main User-Facing Benchmark}
Table~\ref{tab:main-bench} is the main user-facing regression test for the deployed service. The four categories vary along two axes: query complexity (simple vs.\ complex) and answer shape (direct answer vs.\ recommendation). Direct-answer queries ask for a specific fact or a synthesis across catalog resources; recommendation queries ask for catalog items matching a topical, geographic, or multi-criteria filter.

Table~\ref{tab:main-bench} reports the 170 canonical rows for
which all compared methods have aligned outputs and judge scores. I-GUIDE Smart Search leads on overall correctness and remains essentially tied with the best overall faithfulness and completeness scores while dominating the simple categories on correctness and completeness, where answer quality depends on retrieving the correct domain records as represented in the current I-GUIDE index, rather than relying on partial or stale parametric knowledge. 
On simple direct-answer queries, Smart Search reaches high correctness because catalog records supply exact facts such as authors, document specific facts, producer, date, and resource identity; non-retrieval baselines often identify the general topic but invent unsupported details. On simple recommendation, Smart Search's multimodal retrieval surfaces the correct catalog entries, producing a clear correctness gap over non-retrieval baselines.

The complex categories reveal a different tradeoff. Frontier non-retrieval models sometimes match or exceed Smart Search on correctness and completeness for conceptual questions whose source publications are likely represented in pretraining data. 
We treat these rows as a reminder that deployed RAG does not need to win every conceptual question to be valuable; its role is to make answers traceable to the current platform evidence. 
Fabrication instead surfaces on factual queries whose answers require specific source records: on simple direct-answer questions, GPT-4o and GPT-5.5 fall to faithfulness near 20 because they invent producers and posting dates for resources such as government-issued flood risk maps, whereas Smart Search stays grounded in the indexed records.
Naive RAG exhibits the opposite failure mode on complex direct-answer queries: it often refuses or says too little when single-pass semantic retrieval cannot assemble enough evidence. These refusal-style answers contain few unsupported claims, so faithfulness can be high (88 on complex direct-answer) even when correctness and completeness are low (near 31). This is why we read faithfulness, correctness, and completeness together rather than treating any single judge metric as sufficient.

The two answer shapes also expose what faithfulness actually measures. On recommendation queries, a non-retrieval model can answer honestly from partial public or domain knowledge---for example, GPT-5.5 scores faithfulness 96 on simple recommendation by describing FEMA flood mapping without inventing a specific record---yet its correctness stays at 60 because it often fails to identify the exact matching. On direct-answer queries, the model must commit to a specific number, date, producer, or term associated with a knowledge element; when that detail is missing, stale, or only partially represented in parametric knowledge, it may fabricate, and faithfulness collapses to 20. High faithfulness therefore signals only the absence of unsupported claims, and completeness is likewise permissive when it rewards thorough but insufficiently grounded prose. Correctness is the metric that exposes whether the answer identifies the right indexed record and its specific evidence, which is where Smart Search separates from the frontier baselines.


To check whether the aggregate results are driven by easy rows, we also stratify the canonical benchmark by the maximum correctness achieved by any non-retrieval LLM baseline. This separates retrieval-needed rows, where no LLM-only baseline answers well; mixed rows, where LLMs recover partial knowledge but miss detailed facts; and LLM-answerable rows, where frontier models can answer from parametric knowledge. The pattern matches the design purpose of Smart Search: it dominates the retrieval-needed and mixed strata, while Claude Opus 4.7 leads on the LLM-answerable stratum. For an experience paper, this stratification is more important than the aggregate win: it identifies the slice of production traffic for which retrieval architecture, not model scale alone, changes the user outcome. The result shows that Smart Search's gains are concentrated on rows where current I-GUIDE metadata, spatial footprints, graph relationships, and domain-specific records matter, not on questions already covered by frontier LLM training data. Full stratified results are reported in Appendix~\ref{app:difficulty-strata}.

\subsection{Targeted Ablation Study}
\label{sec:ablations}

The Smart Search pipeline was not designed in one pass. Each retrieval and
reasoning component was added in response to a feature request or a
user-facing failure mode observed during deployment. The ablation is therefore written as an engineering audit of accumulated complexity: it asks which pieces remained load-bearing after the system stabilized, and which pieces mainly changed answer quality, coverage, or ranking side effects. The diagonal ablation
therefore asks whether each component still earns its place once the other
components are present.
To make the question tractable we constructed ten intent-specific probe sets with 262 rows total. 
Each set targets one retrieval or generation behavior: exact keyword matching,
spatial filtering, semantic paraphrase resolution, graph traversal, type
filtering, click-popularity ranking, hybrid-modality fusion, multi-hop
reasoning, hallucination resistance, or iterative conversational follow-up.
For each component, we disabled that component and reran the targeted set
with all other components and the generation prompt unchanged. We report
Recall@10 and changes in reference-grounded correctness and completeness on
the same 0--100 scale as Table~\ref{tab:main-bench}. We read these deltas diagnostically: a lower recall score identifies missing evidence, while correctness and completeness show whether the missing evidence changes the answer's accuracy or detail.
We omit faithfulness from this table because, under the honesty rubric, ablating a retrieval component pushes the weakened pipeline toward abstention---which scores as faithful---so $\Delta$F is positive under nearly every ablation and would misleadingly reward component removal; correctness and completeness track the load-bearing contribution instead.

\begin{table*}[t]
\centering
\caption{Diagonal ablation, grouped by the engineering verdict the
table delivers on each module. \emph{Base} and \emph{no-X} are
mean Recall@10 (0--100) for the full Smart Search and the ablated
configuration on the targeted intent set; $\Delta$R@10, $\Delta$C,
and $\Delta$Cp are ablation minus baseline on the same 0--100 scale.
}
\label{tab:ablation-real}
\small
\setlength{\tabcolsep}{5pt}
\begin{tabular}{@{}llcccccc@{}}
\toprule
\textbf{Removed component} & \textbf{Targeted QA set} & \textbf{$n$} & \textbf{Base} & \textbf{no-X} & \textbf{$\Delta$R@10} & \textbf{$\Delta$C} & \textbf{$\Delta$Cp} \\
\midrule
\multicolumn{8}{@{}l}{\emph{Essential modalities}}\\
keyword search   & keyword\_exact       & 30 & 85 & 17           & $-68$ & $-34$ & $-39$  \\
spatial search  & spatial\_required    & 30 & 50 & \phantom{0}7 & $-44$ & $-29$ & $-18$ \\
\midrule
\multicolumn{8}{@{}l}{\emph{Measurable contributors}}\\
semantic search    & semantic\_paraphrase & 28 & 50 & 52 & $+2$  & $-9$ & $0$ \\
RRF reranker        & spatial\_required    & 30 & 50 & 41 & $-9$  & $+2$ & $-1$ \\
memory rewrite      & iterative\_followup  & 30 & 54 & 44 & $-10$ & $-15$ & $-9$ \\
hallucination and relevance velidator & hallucination\_bait  & 29 & \multicolumn{2}{c}{gold-empty}  & --- & $-6$ & $-2$ \\
LLM grader (type)   & type\_filtered       & 30 & 83 & 73 & $-10$ & $-3$  & $-1$  \\
graph search     & graph\_required      & 30 & 28 & 20 & $-8$  & $0$  & $-13$ \\
\midrule
\multicolumn{8}{@{}l}{\emph{Less effective on its target probe set}}\\
iterative reasoning      & multi\_hop        & 26 & 55 & 39 & $-14$ & $+25$ & $+12$ \\
log-popularity boost & popularity\_trap & 29 & 12 & 24 & $+12$ & $+18$ & $+12$ \\
\bottomrule
\end{tabular}

\smallskip
\footnotesize
\end{table*}

Table~\ref{tab:ablation-real} groups the results into three engineering
verdicts. 
Keyword search and spatial search are essential modalities:
removing keyword search drops Recall@10 by 68 points on exact-identifier
queries, and removing the spatial search drops Recall@10 by 44 points on
spatially constrained queries. These rows confirm that exact names and
geographic extent are not reliably recoverable from semantic similarity
alone.


The middle group shows conditional but measurable contributions. Removing semantic search lowers correctness even though Recall@10 is nearly unchanged, suggesting that keyword search can retrieve a similar candidate set but semantic search improves the ordering by using embedded descriptions and document chunks. On \textsc{spatial\_required}, the spatial search has already isolated the correct candidates through geo-filtering, so removing the RRF reranker mostly reorders an already relevant set, costing borderline Recall@10 but not answer quality. Removing memory rewrite lowers both correctness and completeness on iterative follow-up queries. Removing the hallucination and relevance validators lets the pipeline answer hallucination-bait queries it would otherwise refuse, lowering correctness by 6 points. For type grading with the LLM, wrong-type documents enter the top-10 when the grader is removed, but the generator can still filter them at read time when enough correct-type evidence remains. Finally, removing graph search leaves correctness unchanged but reduces completeness by 13 points: the anchor record is usually still recoverable by keyword or semantic retrieval in the I-GUIDE knowledge base, but the weights and relational neighbors that graph edges supply, such as co-contributors and shared bookmarks, are not, so completeness drops.

Two components are less effective on their targeted probes. The iterative loop helps on genuinely complex, multi-hop questions, but on the multi-hop probe it over-iterates on simpler sub-queries---repeatedly digging into the retrieved documents or rephrasing the question without terminating---a behavior we traced to the 7B generator's reasoning limits that persisted even after explicit prompt revision; disabling it raises correctness by $+25$ and completeness by $+12$ at a recall cost of $-14$. The log-popularity boost (Figure~\ref{fig:rerank}) for tie breaking is the other component that underperforms on its
targeted adversarial probe. The popularity-trap set was constructed so that
click frequency points to the wrong answer; disabling the boost therefore
improves the metrics on that slice. We retain the feature because click
signals help on production recommendation traffic, but the ablation shows
that the signal should be down-weighted. This is a typical deployed-system tradeoff rather than a clean algorithmic failure: a UI-derived signal can improve discovery when popularity tracks resource quality, but it needs guardrails when popularity conflicts with topical or spatial relevance.

\subsection{Design Evolution During Evaluation}
Across the 17-month production deployment (December~2024 to May~2026), the
architecture in Section~4 was not built in a single step. It accumulated through
rounds of revision, each triggered by patterns in user queries that the originally
deployed pipeline handled poorly and confirmed afterwards by the benchmark and
ablation evidence above.
Three insufficiencies in the earlier system motivated the major revisions measured above.

First, a text-only LLM grader could not preserve modality-specific evidence
from spatial or graph retrieval because the reason for retrieval often lives
in geometry or edges rather than document text. This motivated
modality-aware RRF over per-method rank lists. Second, open-ended
text-to-Cypher generation was unstable under the small production model:
without examples, Qwen2.5-7B-Instruct often produced invalid or
schema-mismatched Cypher; with unconstrained examples, it overfit to the
demonstrated query shape. This motivated the intent-classified Cypher
catalog with bounded few-shot examples. Third, long prompts occasionally
exceeded the reliable context behavior of the 7B model and produced empty
responses without raising an exception. This motivated a prompt-size
contract that caps candidate documents per grader call, trims schema text
injected into Cypher generation, and logs empty LLM responses as visible
failures.

The ten intent-specific probe sets were added after these revisions to make
the deployment feedback auditable. Each probe set is constructed against the
deployed OpenSearch indexes and Neo4j graph, so its gold evidence is
traceable to the live retrieval surface while isolating behavior that the
original user-facing benchmark did not test.


\subsection{Evaluation Summary}
Overall, the evaluation answers the deployment questions in operational terms. First, the single-A100 experimental stack supports interactive use at small- to medium-gateway scale but exposes GPU saturation beyond roughly 100 concurrent simulated users. Second, the main benchmark shows that Smart Search gains most on platform-grounded queries requiring current I-GUIDE records, spatial footprints, graph relationships, or usage-derived signals. Third, frontier non-retrieval LLMs remain strong on questions already represented in public literature, but lose faithfulness when they supply made-up details. 
Fourth, the ablations show that no single component explains the full gain: keyword and spatial retrieval are load-bearing on their target cases, while semantic retrieval, RRF, memory rewrite, validation, type grading, and graph traversal contribute more conditionally through coverage, correctness, completeness, or refusal behavior; the iterative loop and log-popularity boost are less effective on their target probes because they can add over-iteration or ranking noise.
Fifth, the design-evolution analysis shows why these components entered the deployed system in the first place: modality-blind grading, unstable text-to-Cypher generation, and long-prompt failures under the small model each surfaced as operational failures before becoming architectural revisions.

\section{Experience Lessons and Discussion}
\label{sec:lessons}
The 17-month deployment yielded lessons in three registers: what multimodal
spatial retrieval must represent, what operational disciplines kept the system
honest, and what production scale and the user-facing surface forced on the
design. Each lesson is anchored in Section~5 but describes a system change
motivated by deployment failures, not only a benchmark result.

\subsection{What Multimodal Retrieval and Reasoning Required}

\textbf{Lesson 1: spatial-temporal retrieval is a first-class modality, not a
post-filter.}
Embedding retrieval handles topical similarity, but it does not reliably
recover resources whose relevance comes from geographic containment, neighboring
regions, place-name equivalence, or spatial footprints absent from the text. In
production, location-bearing queries returned topically relevant but
geographically incomplete records. The ablation confirms the pattern: removing the spatial geo-filter on the
\texttt{spatial\_required} probe drops Recall@10 by 44 points and correctness by 29. For Smart Search, spatial retrieval therefore cannot be a
minor post-filter after semantic search; it has to be a routed retrieval path
whose evidence can enter ranking and generation.

\textbf{Lesson 2: graph traversal answers questions text retrieval cannot.}
The Neo4j component became valuable because users asked which resources a
contributor had submitted, which notebooks used a dataset, which publications
were connected to an analysis, and which items were most viewed or reused.
These queries require relationships and usage-derived signals, not only
document text. The graph path also turned \texttt{click\_count}, originally a
frontend sorting field, into a retrieval signal. The \texttt{popularity\_trap}
ablation shows the risk: when click frequency points to the wrong answer, the
signal must be down-weighted rather than trusted blindly.

\textbf{Lesson 3: route iterative reasoning by intent; do not apply it by
default.}
Iterative reasoning helps when a question requires decomposition, multi-hop
traversal, or evidence from several retrieval paths, but it is too expensive as
a default. Complex queries can trigger 20--50 LLM calls and saturate a single
A100 deployment beyond roughly 100 concurrent simulated users. The architecture
therefore keeps short routes for simple keyword, semantic, or spatial lookup
and reserves bounded reasoning, memory rewrite, grading, and validation for
multi-hop, provenance-oriented, or ambiguous questions.

\subsection{Disciplines That Kept the System Honest}

\textbf{Lesson 4: production correctness depends on interface contracts
between components.}
Several deployment failures were contract failures between components rather
than isolated algorithmic failures: a geocoder silently degraded to semantic
fallback, routing under-triggered graph queries, raw click counts overpowered
normalized relevance, and Cypher templates referenced missing schema fields.
Because each failure produced plausible output, the revised system makes
contracts explicit: retrieval paths preserve per-method ranks, RRF avoids
score-scale incompatibility, Cypher generation is bounded by an
intent-classified catalog, and empty model outputs are logged. Correctness
lives as much in these interfaces as in the individual retrieval algorithms.

\textbf{Lesson 5: faithfulness, correctness, and completeness must be read
together.}
Single-metric evaluation hid important failure modes during development.
Surface-similarity scores could not distinguish grounded answers from fluent
fabrication, or refusal from correct abstention. The F/C/Cp triad exposed the
difference: Naive RAG can be faithful on complex direct-answer rows because it
says little, while non-retrieval LLMs often answer fluently but lose
faithfulness when they invent dates, identifiers, or provenance links.
The Table~\ref{tab:ablation-real} update applies the same discipline without using faithfulness as the ablation column: correctness and completeness deltas must still be read against Recall@10, because a higher answer-quality score with lower evidence coverage can indicate over-iteration, shorter answers, or a target-probe artifact rather than a generally better architecture.

\textbf{Lesson 6: refusal is a design choice, not merely an error.}
Smart Search is deliberately scoped to the indexed I-GUIDE knowledge base:
there is no web-search or parametric-knowledge fallback. Out-of-scope questions
may receive an insufficient-information response, but this keeps claims
traceable to retrieved knowledge elements and prevents the system from becoming
a general-purpose chatbot with weak provenance. Future bridges to trusted
external portals should be accountable retrieval paths with validation, not
uncontrolled fallback channels.

\subsection{What Deployment Forced: Scale and Surface}

\textbf{Lesson 7: for catalog-grounded discovery, architecture can matter
more than model scale.}
The difficulty-stratified analysis shows where the system helps. Frontier LLMs
win some conceptual rows already covered by public literature, while Smart
Search gains most on retrieval-needed and mixed rows that depend on current
I-GUIDE records, spatial footprints, graph relationships, or usage signals.
For catalog-grounded discovery, a well-instrumented retrieval architecture
around a small open model can outperform stronger non-retrieval models on the
rows the platform is meant to serve. The serving benchmark gives the boundary:
this is viable at small- to medium-gateway scale, but GPU budget shapes routing,
context size, and concurrency.

\textbf{Lesson 8: the user interface is part of the retrieval system.} The card-and-stream interface exposes cited knowledge elements, lets users inspect sources, and streams intermediate pipeline states during long-running retrieval. It also closes two feedback loops. Clicks and card interactions become usage-derived graph signals, useful when popularity tracks quality but risky when it amplifies the wrong item, as the \texttt{popularity\_trap} probe shows. Per-answer feedback on relevance, sufficiency, accuracy, clarity, completeness, trust, and comments is not yet used as formal evaluation labels, but it gives future versions a way to connect user experience, retrieval traces, and pipeline revision.

\vspace{-0.75\baselineskip}
\subsection{Limitations and Future Work}
This experience report is bounded by one deployed platform. The
benchmark tests Smart Search against I-GUIDE metadata, graph edges, and retrieval
surfaces rather than serving as a community geospatial RAG benchmark. Its
LLM-generated, human-filtered questions and LLM-as-judge scores support
comparison and failure discovery, and the small human audit calibrates their use
for rank-order interpretation, but all remain sensitive to judge calibration,
answer style, refusal, partial correctness, and unsupported specificity.
Operationally, we still need a full user study, horizontal-scaling tests,
sustained traffic measurements, and monitoring for stale metadata, missing
footprints, broken links, and graph-schema drift. Future work extends Smart
Search from knowledge-element QA toward agentic geospatial analysis, where
notebooks, code, and publications become callable workflow resources requiring
tool verification, sandboxing, access control, and provenance tracking.

\vspace{-0.8em}
\section{Conclusion}

This paper presented I-GUIDE Smart Search as both a production
deployment and an experience report on multimodal geospatial RAG inside a
scientific gateway. Smart Search combines production-maintained OpenSearch
keyword, vector, and spatial indexes with a Neo4j provenance graph and an
iterative RAG pipeline for query augmentation, retrieval routing, evidence
fusion, grounded generation, and validation. It is a platform-grounded discovery
service, not a general-purpose chatbot, and answers I-GUIDE-specific questions
with traceable evidence from current knowledge elements.

The evaluation shows where this architecture matters. Smart
Search improves evidence coverage and judged answer quality most clearly for
recommendation, platform-specific lookup, and retrieval-needed or mixed
questions where general-purpose LLMs lack current catalog metadata, spatial
footprints, graph relationships, or usage signals. Frontier non-retrieval LLMs
remain strong on questions already represented in public literature, reinforcing
the main experience lesson: deployed spatial RAG quality depends less on model
scale alone than on maintained spatial metadata, graph provenance, retrieval
traces, validation, refusal discipline, feedback, and cost controls.

 \begin{acks}
Generative AI tools are used for language polishing and editorial assistance; all manuscript content was reviewed and approved by the authors.
\end{acks}

\bibliographystyle{ACM-Reference-Format}
\bibliography{sigspatial2026_submission/references}

\newpage
\appendix
\section{Artifact availability.}
The evaluation artifacts, benchmark queries, scripts, and reproduction
instructions are available at
\url{https://github.com/I-GUIDE/I-GUIDE_Smart_Search_sigspatial26_artifact}.

\section{Cypher Intent Catalog}
\label{app:cypher-catalog}

Table~\ref{tab:cypher-catalog} lists the ten Cypher templates referenced
from Section~\ref{sec:architecture} (architecture) and Section~\ref{sec:lessons}
(Lesson~2). Each template returns the same standard projection so the
downstream reranker and generation stages see a uniform document shape
regardless of which intent fired.

\begin{table}[h]
\caption{Cypher intent catalog used by the graph retrieval path. Each
intent maps to a template that returns the same standard projection,
so the reranker sees a uniform document shape regardless of which
intent fired.}
\label{tab:cypher-catalog}
\small
\setlength{\tabcolsep}{4pt}
\begin{tabular}{@{}p{0.28\columnwidth}p{0.64\columnwidth}@{}}
\toprule
\textbf{Intent} & \textbf{Template shape (sketch)} \\
\midrule
POPULARITY        & order non-Contributor nodes by \texttt{click\_count} desc \\
\midrule
AUTHORSHIP        & match by contributor name, return their elements \\
\midrule
PROVENANCE        & anchor by title $\rightarrow$ \texttt{RELATED} / \texttt{ALIAS\_OF} / \texttt{BELONGS\_TO} \\
\midrule
REVERSE           & elements whose edges point \emph{into} the anchor \\
\midrule
CO\_CONTRIBUTION  & anchor $\leftarrow$ Contributor $\rightarrow$ other elements \\
\midrule
CO\_BOOKMARK      & anchor $\leftarrow$ User $\rightarrow$ other elements \\
\midrule
ALIAS\_RESOLUTION & transitive \texttt{ALIAS\_OF*0..3} from anchor \\
\midrule
AGGREGATION       & rank by degree (\texttt{count(r)}) \\
\midrule
ABSENCE           & elements lacking a relationship (anti-pattern) \\
\midrule
HYBRID            & node-property filter $\wedge$ relationship constraint \\
\bottomrule
\end{tabular}
\end{table}

\subsection{Paraphrase Robustness}
\label{app:paraphrase}

To account for the 100 user-facing QA instances not included in the canonical
170-query comparison, we evaluate regenerated paraphrase variants of the
simple direct-answer and simple recommendation sets. Each paraphrase
preserves the reference knowledge elements and underlying intent while
changing the surface wording of the question. The paraphrases were drafted by
Claude Opus 4.7, so we report this analysis separately from the main
comparison to avoid conflating paraphrase robustness with a possible
generator-style advantage for Claude Opus.

\begin{table}[t]
\centering
\caption{Paraphrase robustness on regenerated variants of the simple
direct-answer and simple recommendation sets. Correctness is reported on a
0--100 scale. Scores use the \texttt{gpt-5-mini} judge.}
\label{tab:paraphrase-bench}
\small
\setlength{\tabcolsep}{6pt}
\begin{tabular}{l ccc}
\toprule
\textbf{Method} & \textbf{Simple Direct v2} & \textbf{Simple Rec.\ v2} & \textbf{Overall} \\
\midrule
Qwen2.5 & 48 & 55 & 51 \\
GPT-4o & 64 & 71 & 68 \\
GPT-5.5 & 71 & \textbf{82} & \textbf{77} \\
Claude Opus 4.7 & \textbf{74} & 74 & 74 \\
Naive RAG & 38 & 55 & 46 \\
I-GUIDE Smart Search & 62 & 69 & 66 \\
\bottomrule
\end{tabular}
\end{table}

Table~\ref{tab:paraphrase-bench} shows that paraphrasing changes the ranking:
GPT-5.5 and Claude Opus 4.7 reach 77 and 74 overall correctness, ahead of Smart Search at 66.
Smart Search still retrieves the correct catalog items more reliably than
Naive RAG, with Recall@10 of 70 versus 37, and it remains more faithful among
the retrieval methods. We interpret the Opus result with caution because the
same model drafted the paraphrases; a model whose answers match the style of
its own paraphrased questions may receive an implicit advantage. The result is
therefore useful as a robustness check, but the headline method comparison in
Table~\ref{tab:main-bench} remains the canonical comparison.

For retrieval-grounded behavior, we additionally judged Naive-RAG and
Smart-Search answers against the documents each pipeline actually retrieved.
Smart Search refused less often than Naive RAG (31\% vs. 55\% of rows); among
substantive answers, evidence-grounded faithfulness was 74 for Smart Search
and 71 for Naive RAG. This suggests that the paraphrase cohort's remaining
errors are driven more by retrieval coverage and answer completeness than by
unsupported generation from retrieved evidence.

\subsection{Difficulty Stratification}
\label{app:difficulty-strata}
To further test whether Smart Search's aggregate gains are driven by retrieval-needed rows or by easy cases, we stratify the canonical benchmark by the maximum correctness achieved by any non-retrieval LLM baseline (Qwen2.5, GPT-4o, GPT-5.5, or Claude Opus 4.7). \emph{Stratum A} contains rows with max-LLM correctness $\le 3$, where retrieval is necessary because no LLM-only baseline answers well. \emph{Stratum B} contains rows with max-LLM correctness 4--6, where LLMs have partial knowledge but retrieval is needed to ground platform-specific facts. \emph{Stratum C} contains rows with max-LLM correctness $\ge 7$, where the answer is largely reachable from parametric model knowledge
\begin{table*}[t]
\centering
\caption{Canonical benchmark stratified by LLM-only row difficulty. Rows are
partitioned by the maximum correctness achieved by any non-retrieval LLM
baseline: Stratum A (max-LLM $\le 3$, retrieval-needed; $n=13$), Stratum B
(4--6, mixed; $n=72$), and Stratum C ($\ge 7$, LLM-answerable; $n=85$).
Metrics are reported on a 0--100 scale. Scores use the \texttt{gpt-5-mini} judge; strata are repartitioned under it.}
\label{tab:main-bench-strata}
\small
\setlength{\tabcolsep}{3.6pt}
\begin{tabular}{l cccc cccc cccc}
\toprule
\multirow{2}{*}{\textbf{Method}}
& \multicolumn{4}{c}{\textbf{Stratum A: Retrieval-needed ($n=13$)}}
& \multicolumn{4}{c}{\textbf{Stratum B: Mixed ($n=72$)}}
& \multicolumn{4}{c}{\textbf{Stratum C: LLM-answerable ($n=85$})} \\
\cmidrule(lr){2-5} \cmidrule(lr){6-9} \cmidrule(lr){10-13}
& R@10 & F & C & Cp
& R@10 & F & C & Cp
& R@10 & F & C & Cp \\
\midrule
Qwen2.5         & -- & 44 & 13 & 35 & -- & 70 & 34 & 48 & -- & 73 & 47 & 57 \\
GPT-4o          & -- & 29 & 12 & 37 & -- & 42 & 27 & 63 & -- & 77 & 63 & 63 \\
GPT-5.5         & -- & 46 & 18 & 45 & -- & 41 & 28 & 66 & -- & \textbf{86} & \textbf{83} & \textbf{81} \\
Claude Opus 4.7 & -- & 42 & 24 & 37 & -- & 81 & 41 & 34 & -- & 76 & 73 & 66 \\
Naive RAG       & 33 & \textbf{97} & 36 & 35 & 50 & 81 & 54 & 55 & 26 & 64 & 31 & 30 \\
\textbf{I-GUIDE Smart Search}
                 & \textbf{96} & 89 & \textbf{69} & \textbf{69}
                 & \textbf{82} & \textbf{82} & \textbf{81} & \textbf{82}
                 & \textbf{60} & 72 & 65 & 62 \\
\bottomrule
\end{tabular}
\end{table*}

The stratification separates three behaviors. On retrieval-needed rows the full pipeline closes most of the gap left by single-pass RAG; on mixed rows retrieval grounds the specific I-GUIDE record, date, or footprint that LLMs miss; and on LLM-answerable rows frontier models answer well from parametric knowledge, so Smart Search's value there is traceability rather than raw correctness dominance.

\subsection{Human Calibration of the Judge}
\label{app:human-calib}
Because all answer-quality metrics are produced by an LLM-as-judge, two authors independently scored a stratified audit sample of 50 anonymized answer instances spanning the four user-facing categories and six methods. The sample retains one malformed query as an unanswerable ``bait'' probe, because how a system handles a null or ill-formed request is itself meaningful. The audit used the 0/1/2 rubric in Table~\ref{tab:human-rubric}, refined during calibration and applied independently to faithfulness, correctness, and completeness. This rubric is question-centric and intentionally differs from the judge's reference-grounded definitions, so the comparison tests whether the judge's rankings track human judgment rather than whether the two operationalizations are identical. Table~\ref{tab:human-calib} reports quadratic-weighted Cohen's $\kappa$ and Spearman $\rho$ after binning the judge's 0--10 scores into the same three bands. On correctness and completeness, judge--human agreement exceeds inter-annotator agreement, and rank correlations are moderate to strong; faithfulness is weaker, consistent with its sensitivity to abstention and unverifiable details. We therefore use the judge primarily for comparative ranking and read faithfulness, correctness, and completeness jointly rather than in isolation.

\begin{table}[h]
\centering
\caption{Human annotation rubric. Each axis is scored $0$/$1$/$2$
independently. Annotators graded from the question, reference, and answer
without the intended query-type label.}
\label{tab:human-rubric}
\small
\setlength{\tabcolsep}{4pt}
\begin{tabular}{@{}p{0.24\columnwidth}p{0.70\columnwidth}@{}}
\toprule
\textbf{Axis} & \textbf{Levels (0 / 1 / 2)} \\
\midrule
\textbf{Faithfulness} &
\textbf{0}: fabricates a specific fact---date, author, citation, provenance, or
recommended resource---even if the overall answer is correct;
\textbf{1}: largely honest with a minor unverifiable detail;
\textbf{2}: no fabrication, and an explicit abstention (``I don't have enough
information'') is fully faithful, as is correct knowledge absent from the
reference. Confident specifics that cannot be verified default toward
fabrication. \\
\addlinespace
\textbf{Correctness} &
\textbf{0}: does not answer the asked question, or the answer is wrong;
\textbf{1}: partially answers, answers only partly correctly, or is plausible
but uncertain;
\textbf{2}: correctly answers the question, with credit for correct content
beyond the reference. Judged from the question and answer, using the reference
and annotator knowledge as support; off-topic answers fail even if internally
true. \\
\addlinespace
\textbf{Completeness} &
\textbf{0}: barely addresses the question;
\textbf{1}: right direction but missing information;
\textbf{2}: thoroughly addresses the question. Scored as comprehensiveness of
the response, independent of correctness and of the reference---a thorough but
partly fabricated answer can still score high. \\
\bottomrule
\end{tabular}
\end{table}

\begin{center}
\centering
\captionof{table}{Human calibration of the \texttt{gpt-5-mini} judge ($n=50$). Judge--human compares the binned judge scores with the two-annotator consensus; because the three-point ordinal scale compresses $\kappa$, we also report rank agreement.}
\label{tab:human-calib}
\small
\begin{tabular}{lccc}
\toprule
\textbf{Dimension} & \textbf{Inter-annotator $\kappa$} & \textbf{Judge--human $\kappa$} & \textbf{$\rho$} \\
\midrule
Faithfulness & 0.33 & 0.25 & 0.33 \\
Correctness  & 0.32 & 0.52 & 0.61 \\
Completeness & 0.26 & 0.69 & 0.68 \\
\bottomrule
\end{tabular}
\end{center}

\end{document}